# Cancer Risk Messages:
# A Light Bulb Model

Ka C. CHAN [a,1], Ruth F. G. WILLIAMS [b,c], Christopher T. LENARD [c] and Terence M. MILLS [c]

[a] *School of Business, Education, Law and Arts, University of Southern Queensland*
*Springfield Central, QLD 4300, Australia*
[b] *Graduate School of Education, The University of Melbourne*
*VIC 3052, Australia*
[c] *Department of Mathematics and Statistics, La Trobe University*
*Bendigo, VIC 3552, Australia*

**Abstract.** The meaning of public messages such as "One in x people gets cancer" or "One in y people gets cancer by age z" can be improved. One assumption commonly invoked is that there is no other cause of death, a confusing assumption. We develop a light bulb model to clarify cumulative risk and we use Markov chain modeling, incorporating the assumption widely in place, to evaluate transition probabilities. Age-progression in the cancer risk is then reported on Australian data. Future modelling can elicit realistic assumptions.

**Introduction**

Messages about the risk of cancer are widespread in many countries. For example, in Australia, "By the age of 85, the risk is estimated to increase to 1 in 2 for males and 1 in 3 for females" [1]. Cancer risk messages are also stated on cancer websites internationally, for instance, the Cancer Australia website [2].

Various questions can come to mind when a risk of cancer message is heard, such as "What does '1 in 2 will get cancer' mean?" and "What does '1 in 2 will get cancer' mean for me?" There are multiple informational dimensions over a population that are condensed to a single reported value about the risk of cancer, such as "1 in 2 men" or "1 in 3 women" by age 85. The meaning drawn from such messages affects not only individuals, but also public health and also economic welfare [3]. Other questions that may be asked, such as "What is the mathematical derivation of this risk?" or "How is such a measure calculated?" are pertinent, and answered fully [4, 5].

Hidden in these cancer risk messages is an assumption of significance. It is vital to be aware that the concept of cumulative incidence risk is based on the assumption that cancer is the only cause of death: "The cumulative risk is the risk an individual would have of developing the disease in question during a certain age period if no other cause of death were in operation" [6]. This paper aims to unravel this confusing implicit assumption that underpins the common approach to risk of cancer calculations. This involves a thought experiment, which we refer to as the light bulb model. We model the risk of cancer with a Markov chain approach to show that the transition probabilities in cancer risk with age-progression and initial Australian results on those transition probabilities.

---

[1] Corresponding Author: Dr K.C. Chan; E-mail: kc.chan@usq.edu.au.

**1. The Light Bulb Model**

For the purpose of this article, it is sufficient to state that the risk of cancer in a population is derived from available data that enumerates cancer incidence and population. These points are illustrated by Table 1.

**Table 1.** Data in five-year age groups.

| Group ($i$) | Age | Population | Cancer Incidence |
|---|---|---|---|
| 1 | 0–4 | $n_1$ | $x_1$ |
| 2 | 5–9 | $n_2$ | $x_2$ |
| ⋮ | ⋮ | ⋮ | ⋮ |
| 15 | 70–74 | $n_{15}$ | $x_{15}$ |

The method for calculating the cumulative risk draws also upon calculating the cumulative rate for a particular region [7]. The cumulative incidence rate of cancer by age $5t$ is given by

$$a(5t) \coloneqq 5\left(\sum_{i=1}^{t} \frac{x_i}{n_i}\right) \tag{1}$$

and the cumulative incidence risk of cancer by age $5t$ is given by

$$r(5t) \coloneqq 1 - \exp(-a(5t)) \tag{2}$$

where $t$ is time step. A time step is defined as 5 years. For example, $t = 15$ corresponds to 75 years of age.

The model presented in this section represents the scenario in a community where the only cause of death is cancer. Suppose that we have a population of light bulbs. Time will be denoted by $t$ and measured in years. Initially (at $t = 0$) all light bulbs are off. In each year, $t = 1, 2, 3 \ldots$, any light bulb which is off will, randomly, either remain off until $t + 1$, or turn on and show red and then stay red for all subsequent years $t + 1, t + 2, t + 3 \ldots$ . A light bulb that is off represents a person who has not been diagnosed with cancer; a light bulb that is red represents a person who has been diagnosed with cancer. What is the probability that a light bulb turns red by year 75?

We use a sequence of diagrams to illustrate what happens to the population over time, Figure 1. Each square is a $1 \times 1$ square; and the sizes of areas represent the proportions of population. In the beginning, at time step $t = 0$, the area of OFF is 1. This means that 100% of the population belongs to the set OFF. Therefore, the probability, $\mathbb{P}_{t=0}(\text{OFF}) = 1$, and $\mathbb{P}_{t=0}(\text{RED}) = 0$.

At time $t > 0$, some light bulbs come on and the area RED appears and grows. As light bulbs can be either OFF or RED at any time, the sets OFF and RED are mutually exclusive. Therefore, $\mathbb{P}_t(\text{RED} \cap \text{OFF}) = 0$, and $\mathbb{P}_t(\text{RED}) + \mathbb{P}_t(\text{OFF}) = 1$. The size of the RED region also represents the probability that a light bulb comes on by time step $t$. The people who get cancer (as RED light bulbs) either continue to live (AC – alive with cancer) or die (DC – died of cancer) over time, i.e., AC ⊆ RED ; DC ⊆ RED, and RED = AC ∪ DC and $\mathbb{P}_t(\text{AC} \cap \text{DC}) = 0$. The sizes of regions in each diagram represent the proportions of the states: RED, OFF, AC, and DC.

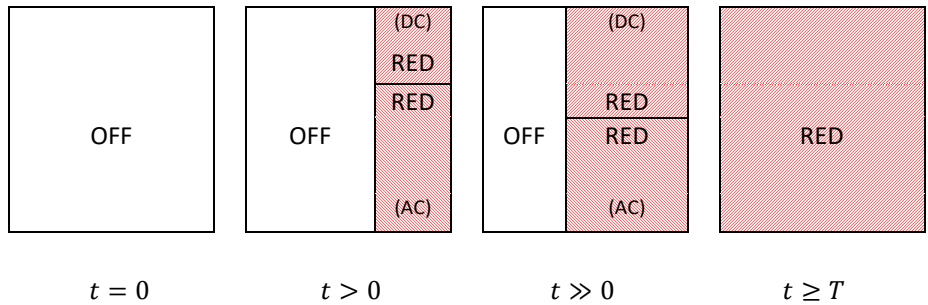

$t = 0$  $\quad\quad\quad$ $t > 0$ $\quad\quad\quad$ $t \gg 0$ $\quad\quad\quad$ $t \geq T$

**Figure 1**. Light bulb model depicted as a sequence of diagrams

The size of the RED region is non-decreasing and OFF is non-increasing over time. The size of the RED region at time step $t$ represents the probability that one is diagnosed of cancer by the age of $5t$, i.e., $\mathbb{P}_t(\text{RED})$. The probability $\mathbb{P}_t(\text{RED}) = \mathbb{P}_t(\text{AC}) + \mathbb{P}_t(\text{DC})$. The size of RED is cumulative in nature, including all light bulbs turned on from time step 0 to $t$. As the whole population reaches a relatively old age, $t = T$, all light bulbs turn RED and 100% of the population belongs to the set RED. Therefore, at time step $t \geq T$, $\mathbb{P}_{t \geq T}(\text{OFF}) = 0$, and $\mathbb{P}_{t \geq T}(\text{RED}) = 1$.

Both cumulative risk, $r(5t)$, and our light bulb model have the same purpose which is to find the sizes of the RED regions in the diagrams, or probabilities of RED, by a certain time. We define two states, S = {OFF, RED} where RED means "diagnosed with cancer" and OFF means "alive and cancer free".

The cumulative risk $r(5t)$ is commonly used as an estimate for the probability $\mathbb{P}_t(\text{Cancer Incidence})$ by time step $t$ or age $5t$. Similarly, the probability $\mathbb{P}_t(\text{RED})$ in our model, is used to estimate $\mathbb{P}_t(\text{Cancer Incidence})$. The state transition diagram for the light bulb model is shown in **Error! Reference source not found.**. The numbers 0 and 1 correspond to the two states {OFF, RED}, and are used as the row or column indices of the transition probability matrix, and the row indices of the state vector.

In the beginning, every newborn baby (100%) in a population is alive and cancer free. Therefore, at time step $t = 0$, $S_0(\text{OFF}) = 1$, and $S_0(\text{RED}) = 0$. The initial state vector is $S_0 = \begin{pmatrix} \text{OFF} & \text{RED} \\ 1 & 0 \end{pmatrix}$. The transition probability matrices are estimated from the observed data, and are time-varying. The four transition probabilities are:

$$\mathbf{P}_i = \begin{bmatrix} P_i(0,0) & P_i(0,1) \\ P_i(1,0) & P_i(1,1) \end{bmatrix} = \begin{bmatrix} P_i(\text{OFF remains OFF}) & P_i(\text{OFF turns RED}) \\ P_i(\text{RED turns OFF}) & P_i(\text{RED remains RED}) \end{bmatrix}.$$

As it is impossible to change from RED to OFF, and RED will stay RED forever (RED is an absorbing state), the probability $P_i(1,0) = 0$, and $P_i(0,1) = 1$. Therefore, we only need to find $P_i(0,0)$ and $P_i(0,1)$ to determine $\mathbf{P}_i$ at each time step, and the transition probability matrix is $\mathbf{P}_i = \begin{bmatrix} P_i(0,0) & P_i(0,1) \\ 0 & 1 \end{bmatrix}$. The state vectors can be determined iteratively: $S_1 = S_0 \mathbf{P}_1$, $S_2 = S_0 \mathbf{P}_1 \mathbf{P}_2$, and $S_t = S_0 \prod_{i=1}^{t} \mathbf{P}_i$. The probability of diagnosed with cancer by time step $t$, or age $5t$, is $\mathbb{P}_t(\text{RED}) := S_t(\text{RED})$.

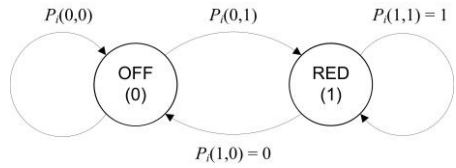
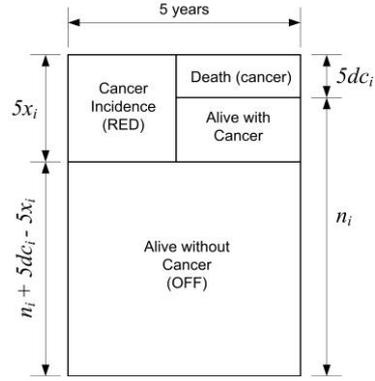

**Figure 2.** State transition diagram for the light bulb model

**Figure 3.** Estimation of transition probabilities

## 2. Estimating Transition Probabilities with Markov Chains

To formulate the light bulb model as a Markov chain, we need to estimate the transition probabilities using the actual data available, similar to that in Table 1. We assume that only living people are counted in the population $n_i$. We also assume that over any time step of five years, the population $n_i$ remains constant, and the total number of people died of cancer would be $5dc_i$, and died of other causes $5do_i$. Therefore, the total number of people in the age group $i$ would be $n_i + 5dc_i + 5do_i$. As the purpose of this model is to simulate cancer incidence rates with the assumption that the only cause of death is cancer, the people who died of other causes are ignored in order to satisfy the assumption requirement. In any age group of people, a person either lives or dies. If they die, they die of cancer. Therefore, the data $do_i$ have been dropped in the estimation of transition probabilities. In the formulation, the total number of people in an age group becomes $n_i + 5dc_i$. Figure 3 shows the state transition of people in an age group over a 5 year period.

The first step is to estimate the number of light bulbs which are OFF at the beginning of a time step, and the numbers of light bulbs remained OFF and turned RED at the end of the same time step. As shown in 0, the estimated number of people who are alive and cancer free (OFF) at the beginning of a time step is $Off_{i,start} = n_i + 5dc_i$. Over this five year period, some people will continue to be cancer free (light bulbs stayed OFF), and some will be diagnosed with cancer (light bulbs turned RED). The number of people who continue to live and cancer free is $Off_{i,end} = n_i + 5(dc_i - x_i)$, and the number of people diagnosed with cancer (light bulbs turned on) is $Red_i = 5x_i$. Therefore, the transition probability $P_i(0,0)$, i.e. OFF remains OFF, is $\frac{Off_{i,end}}{Off_{i,start}} = \frac{n_i + 5(dc_i - x_i)}{n_i + 5dc_i}$.

The transition probability $P_i(0,1)$ can be interpreted as the probability that one would be diagnosed of cancer within the next 5 years. We can estimate $P_i(0,1)$, i.e. from OFF to RED, $P_i(0,1) = \frac{Red_i}{Off_{i,start}} = \frac{5x_i}{n_i+5dc_i}$. Let $B_i = P_i(0,1)$, then $\mathbf{P}_i = \begin{bmatrix} 1-B_i & B_i \\ 0 & 1 \end{bmatrix}$; and $\prod_{i=1}^{t} \mathbf{P}_i = \begin{bmatrix} \prod_{i=1}^{t}(1-B_i) & 1-\prod_{i=1}^{t}(1-B_i) \\ 0 & 1 \end{bmatrix}$. The state vector at time step $t$ is $S_t = [\prod_{i=1}^{t}(1-B_i) \quad 1-\prod_{i=1}^{t}(1-B_i)]$. The probability of diagnosed with cancer by age $5k$ is $\mathbb{P}_t(\text{RED}) := S_t(\text{RED}) = 1 - \prod_{i=1}^{t}\left(1 - \frac{5x_i}{n_i+5dc_i}\right)$.

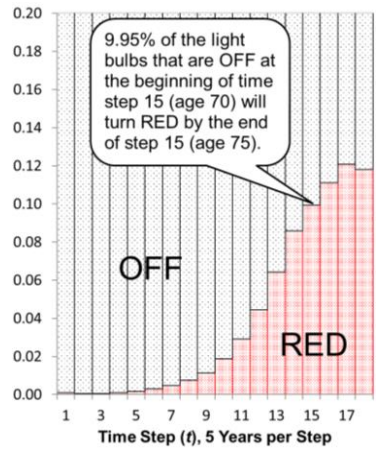 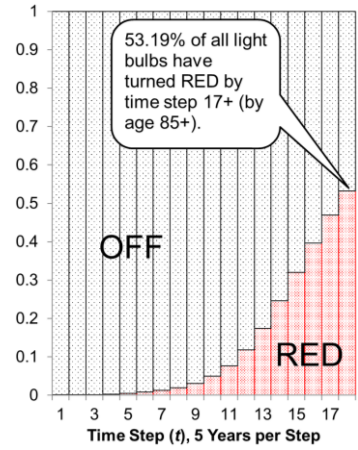

**Figure 4.** Transition probabilities, $P_i(0,1)$

**Figure 5**. Cumulative proportions of RED light bulbs to time step 17+ (age 85+), $\mathbb{P}_t(\text{RED})$

### 3. Results on Australian Data

In this section we report Australian cancer risk in 2010 using the light bulb model with the data for the calculations of cumulative rate, Eq. (1), and cumulative risk, Eq. (2) with data that are publicly available from the AIHW (Australian Institute of Health and Welfare) [8]. The transition probabilities, $P_i(0,1)$, that indicate the age-progression in the risk of cancer, are shown in Figure 4. The cumulative risks, $\mathbb{P}_t(\text{RED})$, are shown in Figure 5. The height of the last column (17+ time step or 85+ age group) shows the value of $\mathbb{P}_{17+}(\text{RED}) = 0.5319$, which is a close estimate of the cumulative risk of 0.5326. This value equates to the "1 in 2" value reported by in the current public health message, although some of the meaning of the signal seems to have been lost by the message. Figure 6 summaries the cumulative rate, cumulative risk, $\mathbb{P}_t(\text{RED})$, and $\mathbb{P}_t(\text{OFF})$.

### 4. Conclusions

Since the concept of cumulative risk behind the "1 in 2" and "1 in 3" messages (regarding Australia) assumes cancer is the only cause of death and since the reported risk of cancer in Australia has encompassed this implicit assumption, the risk in this message is an over-statement. With this assumption invoked, we show how transition probabilities on the age progression in cancer risk can be empirically calculated. Our initial results range widely from a minimum of 0.0011 at the start of life to 0.1182 over 85 years of age. The age progression in cancer risk is apparent in that the probability for the 10–14 year old age group is 0.0006; thirty years on, it is 0.0113 among the 40–44 year old people; and three more decades on, it is 0.0995 in the 70–74 age group.

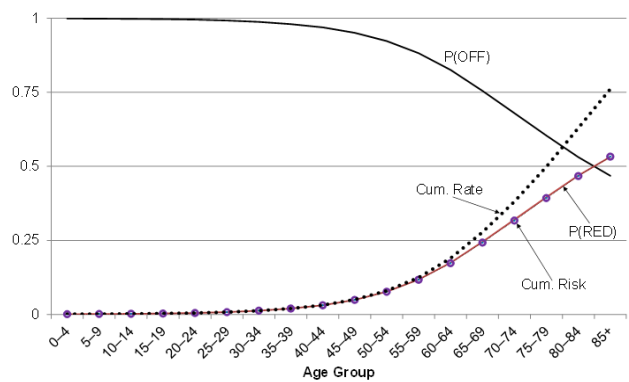

**Figure 6.** RED and OFF probabilities, cumulative risk, and cumulative rate

The model can be developed further to answer many useful questions. More appropriate assumptions can be invoked about causes of death. Also, further applications can compare two regions, or determine the risk of being diagnosed with cancer by a certain age, or of dying from cancer by a particular age, the risk of dying of other causes by a certain age, the chance of being alive with cancer history by a certain age, e.g. "I have no cancer now, what is my chance of cancer in the next 1, 2, 3, 5, 10... years?" or "I had cancer for two years, what is my chance of …?" There is yet more work to be done: Markov chain modeling of the light bulb notion of the cumulative risk of cancer can be used to estimate various aspects of the risk of cancer that enable more meaningful and useful content for reporting to a population on many aspects of the risk of cancer.